 %%%%%%%%%%%%%%%%%%  tex macros for preprints, cm version %%%%%%%%%%%%%%
%                     (P. Ginsparg, last updated 9/91)
%                if confused, type `b' in response to query
%
%---------------------------------------------------------------------%
%% site dependent options:
%% \unredoffs and \redoffs define horizontal and vertical offsets
%% respectively for unreduced and reduced modes. \speclscape defines
%% the \special{} call that sets printer to landscape (sideways) mode.
%% from standard set below, leave uncommented as appropriate or redefine
%
%%% next 400dpi
%\def\unredoffs{} \def\redoffs{\voffset=-.31truein\hoffset=-.48truein}
%\def\speclscape{\special{landscape}}
%
%%% apple lw
\def\unredoffs{} \def\redoffs{\voffset=-.31truein\hoffset=-.59truein}
\def\speclscape{\special{ps: landscape}}
%
%%% qms lasergrafix:
%\def\unredoffs{} \def\redoffs{\voffset=-.4truein\hoffset=.125truein}
%\def\speclscape{\special{qms: landscape}}
%
%%% saclay A4 paper:
%\def\unredoffs{\hoffset-.14truein\voffset-.2truein}
%\def\redoffs{\voffset=-.55truein\hoffset=-.1truein} \def\speclscape{}
%
%---------------------------------------------------------------------%
%
\newbox\leftpage \newdimen\fullhsize \newdimen\hstitle \newdimen\hsbody
\tolerance=1000\hfuzz=2pt
\catcode`\@=11 % This allows us to modify PLAIN macros.
\def\bigans{b }
%\message{ big or little (b/l)? }\read-1 to\answ
\def\answ{b }
\ifx\answ\bigans\message{(This will come out unreduced.}
\magnification=1200\unredoffs\baselineskip=16pt plus 2pt minus 1pt
\hsbody=\hsize \hstitle=\hsize %take default values for unreduced format
\else\message{(This will be reduced.} \let\l@r=L
\magnification=1000\baselineskip=16pt plus 2pt minus 1pt \vsize=7truein
\redoffs \hstitle=8truein\hsbody=4.75truein\fullhsize=10truein\hsize=\hsbody
\output={\ifnum\pageno=0 %%% This is the HUTP version
  \shipout\vbox{\speclscape{\hsize\fullhsize\makeheadline}
    \hbox to \fullhsize{\hfill\pagebody\hfill}}\advancepageno
  \else
  \almostshipout{\leftline{\vbox{\pagebody\makefootline}}}\advancepageno
  \fi}
\def\almostshipout#1{\if L\l@r \count1=1 \message{[\the\count0.\the\count1]}
      \global\setbox\leftpage=#1 \global\let\l@r=R
 \else \count1=2
  \shipout\vbox{\speclscape{\hsize\fullhsize\makeheadline}
      \hbox to\fullhsize{\box\leftpage\hfil#1}}  \global\let\l@r=L\fi}
\fi
%---------------------------------------------------------------------
%
\newcount\yearltd\yearltd=\year\advance\yearltd by -1900

\def\Title#1#2{\nopagenumbers\abstractfont\hsize=\hstitle\rightline{#1}%
\vskip 1in\centerline{\titlefont #2}\abstractfont\vskip .5in\pageno=0}
\def\Date#1{\vfill\leftline{#1}\tenpoint\supereject\global\hsize=\hsbody%
\footline={\hss\tenrm\folio\hss}}%      restores pagenumbers
%
%       use following instead of \Date on the preliminary draft,
%       puts date/time on each page in big mode, writes labels in margins

\def\draftmode{\message{ DRAFTMODE }\def\draftdate{{\rm preliminary draft:
\number\month/\number\day/\number\yearltd\ \ \hourmin}}%
\headline={\hfil\draftdate}\writelabels\baselineskip=20pt plus 2pt minus 2pt
 {\count255=\time\divide\count255 by 60 \xdef\hourmin{\number\count255}
  \multiply\count255 by-60\advance\count255 by\time
  \xdef\hourmin{\hourmin:\ifnum\count255<10 0\fi\the\count255}}}
%       use \nolabels to get rid of eqn, ref, and fig labels in draft mode
\def\nolabels{\def\wrlabeL##1{}\def\eqlabeL##1{}\def\reflabeL##1{}}
\def\writelabels{\def\wrlabeL##1{\leavevmode\vadjust{\rlap{\smash%
{\line{{\escapechar=` \hfill\rlap{\sevenrm\hskip.03in\string##1}}}}}}}%
\def\eqlabeL##1{{\escapechar-1\rlap{\sevenrm\hskip.05in\string##1}}}%
\def\reflabeL##1{\noexpand\llap{\noexpand\sevenrm\string\string\string##1}}}
\nolabels
%
% tagged sec numbers
\global\newcount\secno \global\secno=0
\global\newcount\meqno \global\meqno=1
\def\newsec#1{\global\advance\secno by1\message{(\the\secno. #1)}
%\ifx\answ\bigans \vfill\eject \else \bigbreak\bigskip \fi  %if desired
\global\subsecno=0\eqnres@t\noindent{\bf\the\secno. #1}
\writetoca{{\secsym} {#1}}\par\nobreak\medskip\nobreak}
\def\eqnres@t{\xdef\secsym{\the\secno.}\global\meqno=1\bigbreak\bigskip}
\def\sequentialequations{\def\eqnres@t{\bigbreak}}\xdef\secsym{}
\global\newcount\subsecno \global\subsecno=0
\def\subsec#1{\global\advance\subsecno by1\message{(\secsym\the\subsecno.
#1)}
\ifnum\lastpenalty>9000\else\bigbreak\fi
\noindent{\it\secsym\the\subsecno. #1}\writetoca{\string\quad
{\secsym\the\subsecno.} {#1}}\par\nobreak\medskip\nobreak}
\def\appendix#1#2{\global\meqno=1\global\subsecno=0\xdef\secsym{\hbox{#1.}}
\bigbreak\bigskip\noindent{\bf Appendix #1. #2}\message{(#1. #2)}
\writetoca{Appendix {#1.} {#2}}\par\nobreak\medskip\nobreak}
%
%       \eqn\label{a+b=c}       gives displayed equation, numbered
%                               consecutively within sections.
%     \eqnn and \eqna define labels in advance (of eqalign?)
%
\def\eqnn#1{\xdef #1{(\secsym\the\meqno)}\writedef{#1\leftbracket#1}%
\global\advance\meqno by1\wrlabeL#1}
\def\eqna#1{\xdef #1##1{\hbox{$(\secsym\the\meqno##1)$}}
\writedef{#1\numbersign1\leftbracket#1{\numbersign1}}%
\global\advance\meqno by1\wrlabeL{#1$\{\}$}}
\def\eqn#1#2{\xdef #1{(\secsym\the\meqno)}\writedef{#1\leftbracket#1}%
\global\advance\meqno by1$$#2\eqno#1\eqlabeL#1$$}
%
%                            footnotes
\newskip\footskip\footskip14pt plus 1pt minus 1pt %sets footnote baselineskip
\def\footnotefont{\ninepoint}\def\f@t#1{\footnotefont #1\@foot}
\def\f@@t{\baselineskip\footskip\bgroup\footnotefont\aftergroup\@foot\let\next}
\setbox\strutbox=\hbox{\vrule height9.5pt depth4.5pt width0pt}
\global\newcount\ftno \global\ftno=0
\def\foot{\global\advance\ftno by1\footnote{$^{\the\ftno}$}}
%
%say \footend to put footnotes at end
%will cause problems if \ref used inside \foot, instead use \nref before
\newwrite\ftfile
\def\footend{\def\foot{\global\advance\ftno by1\chardef\wfile=\ftfile
$^{\the\ftno}$\ifnum\ftno=1\immediate\openout\ftfile=foots.tmp\fi%
\immediate\write\ftfile{\noexpand\smallskip%
\noexpand\item{f\the\ftno:\ }\pctsign}\findarg}%
\def\footatend{\vfill\eject\immediate\closeout\ftfile{\parindent=20pt
\centerline{\bf Footnotes}\nobreak\bigskip\input foots.tmp }}}
\def\footatend{}
%
%     \ref\label{text}
% generates a number, assigns it to \label, generates an entry.
% To list the refs on a separate page,  \listrefs
%
\global\newcount\refno \global\refno=1
\newwrite\rfile
\def\ref{[\the\refno]\nref}
\def\nref#1{\xdef#1{[\the\refno]}\writedef{#1\leftbracket#1}%
\ifnum\refno=1\immediate\openout\rfile=refs.tmp\fi
\global\advance\refno by1\chardef\wfile=\rfile\immediate
\write\rfile{\noexpand\item{#1\ }\reflabeL{#1\hskip.31in}\pctsign}\findarg}
%        horrible hack to sidestep tex \write limitation
\def\findarg#1#{\begingroup\obeylines\newlinechar=`\^^M\pass@rg}
{\obeylines\gdef\pass@rg#1{\writ@line\relax #1^^M\hbox{}^^M}%
\gdef\writ@line#1^^M{\expandafter\toks0\expandafter{\striprel@x #1}%
\edef\next{\the\toks0}\ifx\next\em@rk\let\next=\endgroup\else\ifx\next\empty%
\else\immediate\write\wfile{\the\toks0}\fi\let\next=\writ@line\fi\next\relax}}
\def\striprel@x#1{} \def\em@rk{\hbox{}}
\def\lref{\begingroup\obeylines\lr@f}
\def\lr@f#1#2{\gdef#1{\ref#1{#2}}\endgroup\unskip}
\def\semi{;\hfil\break}
\def\addref#1{\immediate\write\rfile{\noexpand\item{}#1}} %now unnecessary
\def\startrefs#1{\immediate\openout\rfile=refs.tmp\refno=#1}
\def\xref{\expandafter\xr@f}\def\xr@f[#1]{#1}
\def\refs#1{\count255=1[\r@fs #1{\hbox{}}]}
\def\r@fs#1{\ifx\und@fined#1\message{reflabel \string#1 is undefined.}%
\nref#1{need to supply reference \string#1.}\fi%
\vphantom{\hphantom{#1}}\edef\next{#1}\ifx\next\em@rk\def\next{}%
\else\ifx\next#1\ifodd\count255\relax\xref#1\count255=0\fi%
\else#1\count255=1\fi\let\next=\r@fs\fi\next}
%

%
% this is ugly, but moore insists
\newwrite\ffile\global\newcount\figno \global\figno=1
\def\fig{fig.~\the\figno\nfig}
\def\nfig#1{\xdef#1{fig.~\the\figno}%
\writedef{#1\leftbracket fig.\noexpand~\the\figno}%
\ifnum\figno=1\immediate\openout\ffile=figs.tmp\fi\chardef\wfile=\ffile%
\immediate\write\ffile{\noexpand\medskip\noexpand\item{Fig.\ \the\figno. }
\reflabeL{#1\hskip.55in}\pctsign}\global\advance\figno by1\findarg}
\def\xfig{\expandafter\xf@g}\def\xf@g fig.\penalty\@M\ {}
\def\figs#1{figs.~\f@gs #1{\hbox{}}}
\def\f@gs#1{\edef\next{#1}\ifx\next\em@rk\def\next{}\else
\ifx\next#1\xfig #1\else#1\fi\let\next=\f@gs\fi\next}
\newwrite\lfile
{\escapechar-1\xdef\pctsign{\string\%}\xdef\leftbracket{\string\{}
\xdef\rightbracket{\string\}}\xdef\numbersign{\string\#}}

\def\writestop{\def\writestoppt{\immediate\write\lfile{\string\pageno%
\the\pageno\string\startrefs\leftbracket\the\refno\rightbracket%
\string\def\string\secsym\leftbracket\secsym\rightbracket%
\string\secno\the\secno\string\meqno\the\meqno}\immediate\closeout\lfile}}
\def\writestoppt{}\def\writedef#1{}
\def\seclab#1{\xdef #1{\the\secno}\writedef{#1\leftbracket#1}\wrlabeL{#1=#1}}
\def\subseclab#1{\xdef #1{\secsym\the\subsecno}%
\writedef{#1\leftbracket#1}\wrlabeL{#1=#1}}
\newwrite\tfile \def\writetoca#1{}
\def\leaderfill{\leaders\hbox to 1em{\hss.\hss}\hfill}
%        use this to write file with table of contents
\def\writetoc{\immediate\openout\tfile=toc.tmp
   \def\writetoca##1{{\edef\next{\write\tfile{\noindent ##1
   \string\leaderfill {\noexpand\number\pageno} \par}}\next}}}
%       and this lists table of contents on second pass

%
\catcode`\@=12 % at signs are no longer letters
%
%        Unpleasantness in calling in abstract and title fonts
\edef\tfontsize{\ifx\answ\bigans scaled\magstep3\else scaled\magstep4\fi}
\font\titlerm=cmr10 \tfontsize \font\titlerms=cmr7 \tfontsize
\font\titlermss=cmr5 \tfontsize \font\titlei=cmmi10 \tfontsize
\font\titleis=cmmi7 \tfontsize \font\titleiss=cmmi5 \tfontsize
\font\titlesy=cmsy10 \tfontsize \font\titlesys=cmsy7 \tfontsize
\font\titlesyss=cmsy5 \tfontsize \font\titleit=cmti10 \tfontsize
\skewchar\titlei='177 \skewchar\titleis='177 \skewchar\titleiss='177
\skewchar\titlesy='60 \skewchar\titlesys='60 \skewchar\titlesyss='60
\def\titlefont{\def\rm{\fam0\titlerm}% switch to title font
\textfont0=\titlerm \scriptfont0=\titlerms \scriptscriptfont0=\titlermss
\textfont1=\titlei \scriptfont1=\titleis \scriptscriptfont1=\titleiss
\textfont2=\titlesy \scriptfont2=\titlesys \scriptscriptfont2=\titlesyss
\textfont\itfam=\titleit \def\it{\fam\itfam\titleit}\rm}
 \ifx\answ\bigans\else scaled\magstep1\fi
\ifx\answ\bigans\def\abstractfont{\tenpoint}\else
\font\abssl=cmsl10 scaled \magstep1
\font\absrm=cmr10 scaled\magstep1 \font\absrms=cmr7 scaled\magstep1
\font\absrmss=cmr5 scaled\magstep1 \font\absi=cmmi10 scaled\magstep1
\font\absis=cmmi7 scaled\magstep1 \font\absiss=cmmi5 scaled\magstep1
\font\abssy=cmsy10 scaled\magstep1 \font\abssys=cmsy7 scaled\magstep1
\font\abssyss=cmsy5 scaled\magstep1 \font\absbf=cmbx10 scaled\magstep1
\skewchar\absi='177 \skewchar\absis='177 \skewchar\absiss='177
\skewchar\abssy='60 \skewchar\abssys='60 \skewchar\abssyss='60
\def\abstractfont{\def\rm{\fam0\absrm}% switch to abstract font
\textfont0=\absrm \scriptfont0=\absrms \scriptscriptfont0=\absrmss
\textfont1=\absi \scriptfont1=\absis \scriptscriptfont1=\absiss
\textfont2=\abssy \scriptfont2=\abssys \scriptscriptfont2=\abssyss
\textfont\itfam=\bigit \def\it{\fam\itfam\bigit}\def\footnotefont{\tenpoint}%
\textfont\slfam=\abssl \def\sl{\fam\slfam\abssl}%
\textfont\bffam=\absbf \def\bf{\fam\bffam\absbf}\rm}\fi
\def\tenpoint{\def\rm{\fam0\tenrm}% switch back to 10-point type
\textfont0=\tenrm \scriptfont0=\sevenrm \scriptscriptfont0=\fiverm
\textfont1=\teni  \scriptfont1=\seveni  \scriptscriptfont1=\fivei
\textfont2=\tensy \scriptfont2=\sevensy \scriptscriptfont2=\fivesy
\textfont\itfam=\tenit
\def\it{\fam\itfam\tenit}\def\footnotefont{\ninepoint}%
\textfont\bffam=\tenbf \def\bf{\fam\bffam\tenbf}\def\sl{\fam\slfam\tensl}\rm}
\font\ninerm=cmr9 \font\sixrm=cmr6 \font\ninei=cmmi9 \font\sixi=cmmi6
\font\ninesy=cmsy9 \font\sixsy=cmsy6 \font\ninebf=cmbx9
\font\nineit=cmti9 \font\ninesl=cmsl9 \skewchar\ninei='177
\skewchar\sixi='177 \skewchar\ninesy='60 \skewchar\sixsy='60
\def\ninepoint{\def\rm{\fam0\ninerm}% switch to footnote font
\textfont0=\ninerm \scriptfont0=\sixrm \scriptscriptfont0=\fiverm
\textfont1=\ninei \scriptfont1=\sixi \scriptscriptfont1=\fivei
\textfont2=\ninesy \scriptfont2=\sixsy \scriptscriptfont2=\fivesy
\textfont\itfam=\ninei \def\it{\fam\itfam\nineit}\def\sl{\fam\slfam\ninesl}%
\textfont\bffam=\ninebf \def\bf{\fam\bffam\ninebf}\rm}
%
%---------------------------------------------------------------------
%

\hyphenation{anom-aly anom-alies coun-ter-term coun-ter-terms}
\def\inv{^{\raise.15ex\hbox{${\scriptscriptstyle -}$}\kern-.05em 1}}

\def\Dsl{\,\raise.15ex\hbox{/}\mkern-13.5mu D} %this one can be subscripted
\def\dsl{\raise.15ex\hbox{/}\kern-.57em\partial}

\font\bigit=cmti10 scaled \magstep1
 %pound sterling
\def\lspace{\ifx\answ\bigans{}\else\qquad\fi}
\def\lbspace{\ifx\answ\bigans{}\else\hskip-.2in\fi} % $$\lbspace...$$
\def\boxeqn#1{\vcenter{\vbox{\hrule\hbox{\vrule\kern3pt\vbox{\kern3pt
           \hbox{${\displaystyle #1}$}\kern3pt}\kern3pt\vrule}\hrule}}}
\def\mbox#1#2{\vcenter{\hrule \hbox{\vrule height#2in
               \kern#1in \vrule} \hrule}}  %e.g. \mbox{.1}{.1}
%       matters of taste
%\def\tilde{\widetilde} \def\bar{\overline} \def\hat{\widehat}
%
% some sample definitions
  %     curly letters

\def\e#1{{\rm e}^{^{\textstyle#1}}}

\def\darr#1{\raise1.5ex\hbox{$\leftrightarrow$}\mkern-16.5mu #1}
 %pound sterling
\def\ha{{1\over2}}
\def\half{{\textstyle{1\over2}}} %puts a small half in a displayed eqn
\def\roughly#1{\raise.3ex\hbox{$#1$\kern-.75em\lower1ex\hbox{$\sim$}}}

%\input harvmac.tex

%%%%%%%%%%%%%%%%%%%%%  Rublenye bukvy   %%%%%%%%%%%%%%%%%%%%%%%%
\def\IB{\relax\hbox{$\inbar\kern-.3em{\rm B}$}}
\def\IC{\relax\hbox{$\inbar\kern-.3em{\rm C}$}}
\def\ID{\relax\hbox{$\inbar\kern-.3em{\rm D}$}}
\def\IE{\relax\hbox{$\inbar\kern-.3em{\rm E}$}}
\def\IF{\relax\hbox{$\inbar\kern-.3em{\rm F}$}}
\def\IG{\relax\hbox{$\inbar\kern-.3em{\rm G}$}}
\def\IGa{\relax\hbox{${\rm I}\kern-.18em\Gamma$}}
\def\IH{\relax{\rm I\kern-.18em H}}
\def\IK{\relax{\rm I\kern-.18em K}}
\def\II{\relax{\rm I\kern-.18em I}}
\def\IL{\relax{\rm I\kern-.18em L}}
\def\IP{\relax{\rm I\kern-.18em P}}
\def\IR{\relax{\rm I\kern-.18em R}}
\def\IZ{\relax\ifmmode\mathchoice {\hbox{\cmss Z\kern-.4em Z}}{\hbox{\cmss
Z\kern-.4em Z}} {\lower.9pt\hbox{\cmsss Z\kern-.4em Z}}
{\lower1.2pt\hbox{\cmsss Z\kern-.4em Z}}\else{\cmss Z\kern-.4em Z}\fi}

\def\IB{\relax{\rm I\kern-.18em B}}
\def\IC{{\relax\hbox{$\inbar\kern-.3em{\rm C}$}}}
\def\ID{\relax{\rm I\kern-.18em D}}
\def\IE{\relax{\rm I\kern-.18em E}}
\def\IF{\relax{\rm I\kern-.18em F}}

%%%%%%%%%%%%%%%%%%%% Calligraphic letters  %%%%%%%%%%%%%%%%%%%%%%%

%%%%%%%%%%%%%%%%%%%%%%%%%% Derivatives  %%%%%%%%%%%%%%%%%%%%%%%%
\def\p{\partial}

%%Beltrami

%%%%%%%%%%%%%%%%%%%% letters with bar %%%%%%%%%%%%%%%%%%%%%%%%%%

%%%%%%%%%%%%%%%%%%%%%%%%%%% Math symbols %%%%%%%%%%%%%%%%%%%%%%%

%%%%%%%%%%%%%%%%%%%%% Short Cuts %%%%%%%%%%%%%%%%%%%%%%%

\def\half {{1\over 2}}

%%%%%%%%%%%%%%%%%% Greek %%%%%%%%%%%%%%%%%%%%%%

\def\a{\alpha}
\def\b{\beta}
\def\g{\gamma}  
\def\d{\delta}

\def\l{\lambda} 
\def\k{\kappa}
\def\e{\epsilon}

\def\ha{{\hat\alpha}}

\def\hbe{{\hat\beta}}
\def\hg{{\hat\gamma}}  
  
\def\ho{{\hat\omega}}
\def\hht{{\hat \theta}}

\def\hA{{\hat A}}

%%%%%%%%%%%%%%%%%% Big ( )  %%%%%%%%%%%%%%%%%%%
\def\|{\Big|}
\def\({\Big(}   \def\){\Big)}
\def\[{\Big[}   \def\]{\Big]}

%%%%%%%%%%%%%%%%%%% Something to deal with sub-sub-sections
%%%%%%%%%%%%%%%%%%%%%%%%%%%%%%%%%%%%%%%%%

\def\unlockat{\catcode`\@=11}
\def\lockat{\catcode`\@=12}

\unlockat

% Something to deal with sub-sub-sections

\def\newsec#1{\global\advance\secno by1\message{(\the\secno. #1)}
\global\subsecno=0\global\subsubsecno=0\eqnres@t\noindent {\bf\the\secno. #1}
\writetoca{{\secsym} {#1}}\par\nobreak\medskip\nobreak}
\global\newcount\subsecno \global\subsecno=0
\def\subsec#1{\global\advance\subsecno by1\message{(\secsym\the\subsecno.
#1)}
\ifnum\lastpenalty>9000\else\bigbreak\fi\global\subsubsecno=0
\noindent{\it\secsym\the\subsecno. #1}
\writetoca{\string\quad {\secsym\the\subsecno.} {#1}}
\par\nobreak\medskip\nobreak}
\global\newcount\subsubsecno \global\subsubsecno=0
\def\subsubsec#1{\global\advance\subsubsecno by1
\message{(\secsym\the\subsecno.\the\subsubsecno. #1)}
\ifnum\lastpenalty>9000\else\bigbreak\fi
\noindent\quad{\secsym\the\subsecno.\the\subsubsecno.}{#1}
\writetoca{\string\qquad{\secsym\the\subsecno.\the\subsubsecno.}{#1}}
\par\nobreak\medskip\nobreak}

\def\subsubseclab#1{\DefWarn#1\xdef #1{\noexpand\hyperref{}{subsubsection}%
{\secsym\the\subsecno.\the\subsubsecno}%
{\secsym\the\subsecno.\the\subsubsecno}}%
\writedef{#1\leftbracket#1}\wrlabeL{#1=#1}}% Macros for boxes
\lockat

%why???\font\manual=manfnt
\def\dbend{\lower3.5pt\hbox{\manual\char127}}

%%%%%%%%%%%%%%%%%%% Macros for boxes %%%%%%%%%%%%%%%%%%

\def\boxit#1{\vbox{\hrule\hbox{\vrule\kern8pt
\vbox{\hbox{\kern8pt}\hbox{\vbox{#1}}\hbox{\kern8pt}}
\kern8pt\vrule}\hrule}}
\def\mathboxit#1{\vbox{\hrule\hbox{\vrule\kern8pt\vbox{\kern8pt
\hbox{$\displaystyle #1$}\kern8pt}\kern8pt\vrule}\hrule}}

\overfullrule=0pt

%%%%%%%%%%%%%%%%%%%%%%% Derivatives  %%%%%%%%%%%%%%

\def\p{\partial}

%%Beltrami

%%%%%%%%%%%%%%%%%%%% letters with bar %%%%%%%%%%%%%%%%%%%

%%%%%%%%%%%%%%%%%%%%% Short Cuts %%%%%%%%%%%%%%%%%%%%%%%

\def\half {{1\over 2}}

%%%%%%%%%%%%%%%%%% Greek %%%%%%%%%%%%%%%%%%%%%%

\def\a{\alpha}
\def\b{\beta}
\def\g{\gamma}  
\def\d{\delta}

\def\l{\lambda} 
\def\k{\kappa}
\def\e{\epsilon}

\def\a{\alpha}
\def\b{\beta}
\def\d{\delta}

\def\l{\lambda}

\def\k{\kappa}

\def\t{\theta}

%%%%%%%%%%%%%%%%%% Big ( )  %%%%%%%%%%%%%%%%%%%%%%

\def\|{\Big|}
\def\({\Big(}   \def\){\Big)}
\def\[{\Big[}   \def\]{\Big]}

%%%%%%%%%%%%%%%%%% Text %%%%%%%%%%%%%%%%%%%%%%

%%%%%%%%%%%%%%%%%%%%%%%%%%%%%%%%%%%%%%%%%%%%%%%%%%

\Title{\vbox{\hbox{YITP-SB-02-08}
\hbox{hep-th/yymmxxxx}
}} 
{\vbox{ 
\centerline{The Massless Spectrum of}
\vskip .2cm
\centerline{Covariant Superstrings}}} 
\medskip\centerline{P.A. Grassi$^{~a,}$\foot{pgrassi@insti.physics.sunysb.edu}, 
G. Policastro$^{~b,}$\foot{g.policastro@sns.it}, 
and 
P. van Nieuwenhuizen$^{~a,}$\foot{vannieu@insti.physics.sunysb.edu}}
\medskip 
\centerline{$^{(a)}$ {\it C.N. Yang Institute for Theoretical Physics,} }
\centerline{\it State University of New York at Stony Brook, 
NY 11794-3840, USA}
\vskip .3cm
\centerline{$^{(b)}$ {\it Scuola Normale Superiore,} }
\centerline{\it Piazza dei Cavalieri 7, Pisa, 56126, Italy}
\medskip
\vskip  .5cm
\noindent
We obtain the correct cohomology at any ghost number for the open and closed covariant 
superstring, quantized by an approach which we recently developed. We 
define physical states by the usual condition of BRST invariance and a new condition 
involving a new current which is related to a grading of the underlying affine Lie algebra. 

\Date{February 19,  2002}

\lref\lrp{
U.~Lindstr\"om, M.~Ro\v cek, and P.~van Nieuwenhuizen, in preparation. 
}

\lref\pr{
P. van Nieuwenhuizen, in {\it Supergravity `81}, 
Proceedings First School on Supergravity, Cambridge University Press, 1982, page 165.
}

\lref\polc{
J.~Polchinski,
{\it String Theory. Vol. 1: An Introduction To The Bosonic String,}
{\it String Theory. Vol. 2: Superstring Theory And Beyond,}
{\it  Cambridge, UK: Univ. Pr. (1998) 531 p}.
}

\lref\superstring{
M.~B.~Green and  J.~H.~Schwarz, {\it Covariant Description Of Superstrings,} 
Phys.\  Lett.\ {\bf B136} (1984) 367; M.~B.~Green and J.~H.~Schwarz,
  {\it Properties Of The Covariant Formulation Of Superstring Theories,}
  Nucl.\ Phys.\ {\bf B243} (1984) 285\semi
M.~B.~Green and C.~M.~Hull, QMC/PH/89-7
{\it Presented at Texas A and M Mtg. on String Theory, College
  Station, TX, Mar 13-18, 1989}\semi
R.~Kallosh and M.~Rakhmanov, Phys.\ Lett.\  {\bf B209} (1988) 233\semi
%M.~B.~Green and C.~M.~Hull, ``The Brst Cohomology Of An N=1 Superparticle,''
%{\it  In *College Station 1990, Proceedings, Strings 90* 133-147. }; 
U. ~Lindstr\"om, M.~Ro\v cek, W.~Siegel, 
P.~van Nieuwenhuizen and A.~E.~van de Ven, Phys. Lett. {\bf B224} (1989) 
285, Phys. Lett. {\bf B227}(1989) 87, and Phys. Lett. {\bf B228}(1989) 53; 
S.~J.~Gates, M.~T.~Grisaru, U.~Lindstr\"om, M.~Ro\v cek, W.~Siegel, 
P.~van Nieuwenhuizen and A.~E.~van de Ven,
{\it Lorentz Covariant Quantization Of The Heterotic Superstring,}
Phys.\ Lett.\  {\bf B225} (1989) 44; 
A.~Mikovic, M.~Rocek, W.~Siegel, P.~van Nieuwenhuizen, J.~Yamron and
A.~E.~van de Ven, Phys.\ Lett.\  {\bf B235} (1990) 106; 
U.~Lindstr\"om, M.~Ro\v cek, W.~Siegel, P.~van Nieuwenhuizen and
A.~E.~van de Ven, 
{\it Construction Of The Covariantly Quantized Heterotic Superstring,}
Nucl.\ Phys.\  {\bf B330} (1990) 19 \semi
F. Bastianelli, G. W. Delius and E. Laenen, Phys. \ Lett. \ {\bf
  B229}, 223 (1989)\semi
R.~Kallosh, Nucl.\ Phys.\ Proc.\ Suppl.\  {\bf 18B}
  (1990) 180 \semi
M.~B.~Green and C.~M.~Hull, Mod.\ Phys.\ Lett.\  {\bf A5} (1990) 1399\semi 
M.~B.~Green and C.~M.~Hull, Nucl.\ Phys.\  {\bf B344} (1990) 115\semi
F.~Essler, E.~Laenen, W.~Siegel and J.~P.~Yamron, Phys.\ Lett.\  {\bf B254} (1991) 411\semi 
  F.~Essler, M.~Hatsuda, E.~Laenen, W.~Siegel, J.~P.~Yamron, T.~Kimura
  and A.~Mikovic, 
  Nucl.\ Phys.\  {\bf B364} (1991) 67\semi 
J.~L.~Vazquez-Bello,
  Int.\ J.\ Mod.\ Phys.\  {\bf A7} (1992) 4583\semi
E. Bergshoeff, R. Kallosh and A. Van Proeyen, ``Superparticle
  actions and gauge fixings'', Class.\ Quant.\ Grav {\bf 9} 
  (1992) 321\semi
C.~M.~Hull and J.~Vazquez-Bello, Nucl.\ Phys.\  {\bf B416}, (1994) 173 [hep-th/9308022]\semi
P.~A.~Grassi, G.~Policastro and M.~Porrati,
{\it Covariant quantization of the Brink-Schwarz superparticle,}
Nucl.\ Phys.\ B {\bf 606}, 380 (2001)
[arXiv:hep-th/0009239].
}

\lref\bv{
N. Berkovits and C. Vafa,
{\it $N=4$ Topological Strings}, Nucl. Phys. B433 (1995) 123, 
hep-th/9407190.}

\lref\fourreview{N. Berkovits,  {\it Covariant Quantization Of
The Green-Schwarz Superstring In A Calabi-Yau Background,}
Nucl. Phys. {\bf B431} (1994) 258, ``A New Description Of The Superstring,''
Jorge Swieca Summer School 1995, p. 490, hep-th/9604123.}

\lref\oo{
%\lref\OoguriPS{
H.~Ooguri, J.~Rahmfeld, H.~Robins and J.~Tannenhauser,
{\it Holography in superspace,}
JHEP {\bf 0007}, 045 (2000)
[arXiv:hep-th/0007104].
%%CITATION = HEP-TH 0007104;%%
}

\lref\bvw{
N.~Berkovits, C.~Vafa and E.~Witten,
{\it Conformal field theory of AdS background with Ramond-Ramond flux,}
JHEP {\bf 9903}, 018 (1999)
[arXiv:hep-th/9902098].
%%CITATION = HEP-TH 9902098;%%
}
\lref\wittwi{
E.~Witten,
{\it An Interpretation Of Classical Yang-Mills Theory,}
Phys.\ Lett.\ B {\bf 77}, 394 (1978); 
E.~Witten,
{\it Twistor - Like Transform In Ten-Dimensions,}
Nucl.\ Phys.\ B {\bf 266}, 245 (1986); 
J.~P.~Harnad and S.~Shnider,
{\it Constraints And Field Equations For Ten-Dimensional 
Super-Yang-Mills Theory,}
Commun.\ Math.\ Phys.\  {\bf 106}, 183 (1986).
}

\lref\SYM{
W.~Siegel,
{\it Superfields In Higher Dimensional Space-Time,}
Phys.\ Lett.\ B {\bf 80}, 220 (1979)\semi
B.~E.~Nilsson,
{\it Pure Spinors As Auxiliary Fields In The Ten-Dimensional 
Supersymmetric Yang-Mills Theory,}
Class.\ Quant.\ Grav.\  {\bf 3}, L41 (1986); 
B.~E.~Nilsson,
{\it Off-Shell Fields For The Ten-Dimensional Supersymmetric 
Yang-Mills Theory,} GOTEBORG-81-6\semi
S.~J.~Gates and S.~Vashakidze,
{\it On D = 10, N=1 Supersymmetry, Superspace Geometry And Superstring Effects,}
Nucl.\ Phys.\ B {\bf 291}, 172 (1987)\semi
M.~Cederwall, B.~E.~Nilsson and D.~Tsimpis,
{\it The structure of maximally supersymmetric Yang-Mills theory:  
Constraining higher-order corrections,}
JHEP {\bf 0106}, 034 (2001)
[arXiv:hep-th/0102009]; 
M.~Cederwall, B.~E.~Nilsson and D.~Tsimpis,
{\it D = 10 superYang-Mills at O(alpha**2),}
JHEP {\bf 0107}, 042 (2001)
[arXiv:hep-th/0104236].
}
\lref\har{
J.~P.~Harnad and S.~Shnider,
{\it Constraints And Field Equations For Ten-Dimensional 
Super-Yang-Mills Theory,}
Commun.\ Math.\ Phys.\  {\bf 106}, 183 (1986).
}
\lref\wie{
%\lref\WiegmannHN{
P.~B.~Wiegmann,
{\it Multivalued Functionals And Geometrical Approach 
For Quantization Of Relativistic Particles And Strings,} 
Nucl.\ Phys.\ B {\bf 323}, 311 (1989).
%%CITATION = NUPHA,B323,311;%%
}
\lref\purespinors{\'E. Cartan, {\it Lecons sur la th\'eorie des spineurs}, 
Hermann, Paris (1937)\semi
C. Chevalley, {\it The algebraic theory of Spinors}, 
Columbia Univ. Press., New York\semi
 R. Penrose and W. Rindler, 
{\it Spinors and Space-Time}, Cambridge Univ. Press, Cambridge (1984) 
\semi
P. Budinich and A. Trautman, {\it The spinorial chessboard}, Springer, 
New York (1989).
}
\lref\coset{
P.~Furlan and R.~Raczka,
{\it Nonlinear Spinor Representations,}
J.\ Math.\ Phys.\  {\bf 26}, 3021 (1985)\semi
%%CITATION = JMAPA,26,3021;%%
A.~S.~Galperin, P.~S.~Howe and K.~S.~Stelle,
{\it The Superparticle and the Lorentz group,}
Nucl.\ Phys.\ B {\bf 368}, 248 (1992)
[arXiv:hep-th/9201020].
%%CITATION = HEP-TH 9201020;%%
}

%
%\lref\superspace{
%S.~J.~Gates, M.~T.~Grisaru, M.~Rocek and W.~Siegel,
%{\it Superspace, Or One Thousand 
%And One Lessons In Supersymmetry,''}
%Front.\ Phys.\  {\bf 58}, 1 (1983)
%[arXiv:hep-th/0108200].}
%
\lref\GS{M.B. Green, J.H. Schwarz, and E. Witten, {\it Superstring Theory,} 
 vol. 1, chapter 5 (Cambridge U. Press, 1987).  
}
\lref\carlip{S. Carlip, 
{\it Heterotic String Path Integrals with the Green-Schwarz 
Covariant Action}, Nucl. Phys. B284 (1987) 365 \semi R. Kallosh, 
{\it Quantization of Green-Schwarz Superstring}, Phys. Lett. B195 (1987) 369.} 
 \lref\john{G. Gilbert and 
D. Johnston, {\it Equivalence of the Kallosh and Carlip Quantizations 
of the Green-Schwarz Action for the Heterotic String}, Phys. Lett. B205 
(1988) 273.} 
\lref\csm{W. Siegel, {\it Classical Superstring Mechanics}, Nucl. Phys. B263 (1986) 
93\semi 
W.~Siegel, {\it Randomizing the Superstring}, Phys. Rev. D 50 (1994), 2799.
}   
\lref\sok{E. Sokatchev, {\it 
Harmonic Superparticle}, Class. Quant. Grav. 4 (1987) 237\semi 
E.R. Nissimov and S.J. Pacheva, {\it Manifestly Super-Poincar\'e 
Covariant Quantization of the Green-Schwarz Superstring}, 
Phys. Lett. B202 (1988) 325\semi 
R. Kallosh and M. Rakhmanov, {\it Covariant Quantization of the 
Green-Schwarz Superstring}, Phys. Lett. B209 (1988) 233.}  
\lref\many{S.J. Gates Jr, M.T. Grisaru, 
U. Lindstrom, M. Rocek, W. Siegel, P. van Nieuwenhuizen and 
A.E. van de Ven, {\it Lorentz-Covariant Quantization of the Heterotic 
Superstring}, Phys. Lett. B225 (1989) 44\semi 
R.E. Kallosh, {\it Covariant Quantization of Type IIA,B 
Green-Schwarz Superstring}, Phys. Lett. B225 (1989) 49\semi 
M.B. Green and C.M. Hull, {\it Covariant Quantum Mechanics of the 
Superstring}, Phys. Lett. B225 (1989) 57.}  
 \lref\fms{D. Friedan, E. Martinec and S. Shenker, 
{\it Conformal Invariance, Supersymmetry and String Theory}, 
Nucl. Phys. B271 (1986) 93.}
\lref\kawai{
T.~Kawai,
{\it Remarks On A Class Of BRST Operators,}
Phys.\ Lett.\ B {\bf 168}, 355 (1986).}
 \lref\ufive{N. Berkovits, {\it 
Quantization of the Superstring with Manifest U(5) Super-Poincar\'e 
Invariance}, Phys. Lett. B457 (1999) 94, hep-th/9902099.}  
\lref\BerkovitsRB{ N.~Berkovits, 
{\it Covariant quantization of the superparticle 
using pure spinors,} [hep-th/0105050].  
%%CITATION =HEP-TH 0105050;%% 
} 
%%% berkovits %%%%
%%% berkovits %%%%

\lref\berko{
%\BerkovitsFE
%\lref\BerkovitsFE{
N.~Berkovits,
{\it Super-Poincar\'e covariant quantization of the superstring,}
JHEP { 0004}, 018 (2000)
[hep-th/0001035]%}
\semi
%\BerkovitsPH
%\lref\BerkovitsPH{
N.~Berkovits and B.~C.~Vallilo,
{\it Consistency of super-Poincar\'e covariant superstring tree amplitudes,}
JHEP { 0007}, 015 (2000)
[hep-th/0004171]%}
\semi
%\BerkovitsNN
%\lref\BerkovitsNN{
N.~Berkovits,
{\it Cohomology in the pure spinor formalism for the superstring,}
JHEP { 0009}, 046 (2000)
[hep-th/0006003]%}
\semi
%\BerkovitsWM
%\lref\BerkovitsWM{
N.~Berkovits,
{\it Covariant quantization of the superstring,}
Int.\ J.\ Mod.\ Phys.\ A { 16}, 801 (2001)
[hep-th/0008145]%}
\semi
%\BerkovitsYR
%\lref\BerkovitsYR{
N.~Berkovits and O.~Chandia,
{\it Superstring vertex operators in an AdS(5) x S(5) background,}
Nucl.\ Phys.\ B {\bf 596}, 185 (2001)
[hep-th/0009168]%}
\semi
%\BerkovitsZY
%\lref\BerkovitsZY{
N.~Berkovits,
{\it The Ten-dimensional Green-Schwarz 
superstring is a twisted Neveu-Schwarz-Ramond string,}
Nucl.\ Phys.\ B {\bf 420}, 332 (1994)
[hep-th/9308129]
%%CITATION = HEP-TH 9308129;%%%}
\semi
%\BerkovitsUS
%\lref\BerkovitsUS{
N.~Berkovits,
{\it Relating the RNS and pure spinor formalisms for the superstring,}
[hep-th/0104247]%}
\semi
%\BerkovitsMX
%\lref\BerkovitsMX{
N.~Berkovits and O.~Chandia,
{\it Lorentz invariance of the pure spinor BRST cohomology 
for the  superstring,}
[hep-th/0105149].
}

%\GrassiUG
\lref\GrassiUG{
P.~A.~Grassi, G.~Policastro, M.~Porrati and P.~Van Nieuwenhuizen,
{\it Covariant quantization of superstrings without pure spinor constraints}, 
[hep-th/0112162].
%%CITATION = HEP-TH 0112162;%%
}

\lref\kacmoody{
V.~G.~Kac, {\it Simple graded Lie algebras of finite growth}, Func. Anal. Appl. {\bf 1}, 
(1967) 328; 
K.~Bardacki and M.~B.~Halpern, {\it New dual quark model}, Phys. Rev. D 3 (1971) 2493; 
V.~G.~Kac, {\it Infinite dimensional Lie algebras}, 3rd edition, Cambridge University Press, 
Cambridge, 1990.
}

%\BerkovitsUE
\lref\BerkovitsUE{
N.~Berkovits and P.~Howe,
{\it 
Ten-dimensional supergravity constraints from the pure spinor formalism  
for the superstring}, [hep-th/0112160].
%%CITATION = HEP-TH 0112160;%%
}

%\WittenZZ
\lref\WittenZZ{
E.~Witten,
{\it Mirror manifolds and topological field theory,}
hep-th/9112056.
%%CITATION = HEP-TH 9112056;%%
}

\lref\wichen{
E.~Witten,
{\it Chern-Simons gauge theory as a string theory,}
arXiv:hep-th/9207094.
%%CITATION = HEP-TH 9207094;%%
}

%\lref\bgi{
%C.~Becchi, S.~Giusto and C.~Imbimbo,
%{\it The holomorphic anomaly of topological strings,}
%Fortsch.\ Phys.\  {\bf 47}, 195 (1999)
%[hep-th/9801100]\semi
%C.~Becchi, S.~Giusto and C.~Imbimbo, 
%{\it Topological B models}, unpublished. 
%}

%\howe
\lref\howe{P.S. Howe, {\it Pure Spinor Lines in Superspace and 
Ten-Dimensional Supersymmetric Theories}, 
Phys. Lett. B258 (1991) 141, Addendum-ibid.B259 (1991) 511\semi 
P.S. Howe, {\it Pure Spinors, Function Superspaces and Supergravity 
Theories in Ten Dimensions and Eleven Dimensions}, Phys. Lett. B273 (1991) 
90.}

\lref\equivariant{
R.~Stora,
{\it Exercises in equivariant cohomology,}
CERN-TH-96-279
{\it NATO Advanced Study Institute: Quantum Fields and 
Quantum Space Time, Cargese, Corsica, France, 22 Jul - 3 Aug 1996}.
R.~Stora,
{\it Exercises in equivariant cohomology,} arXiv:hep-th/9611114
\semi J.~Kalkman,
{\it Brst Model For Equivariant Cohomology And Representatives For The Equivariant Thom Class,} 
Commun.\ Math.\ Phys.\  {\bf 153}, 447 (1993)\semi
S.~Ouvry, R.~Stora and P.~van Baal,
{\it On The Algebraic Characterization Of Witten's Topological Yang-Mills Theory,}
Phys.\ Lett.\ B {\bf 220}, 159 (1989)\semi
L.~Baulieu and I.~M.~Singer,
{\it Conformally Invariant Gauge Fixed Actions For 2-D Topological Gravity,}
Commun.\ Math.\ Phys.\  {\bf 135}, 253 (1991).}

\lref\equiva{
P.~A.~Grassi, G.~Policastro, M.~Porrati and P.~Van Nieuwenhuizen, 
{\it  On the BRST Cohomology of Superstrings 
with/without Pure Spinors,} in preparation. 
}

\lref\brscohotopo{
A.~Blasi and R.~Collina,
{\it Basic Cohomology Of Topological Quantum Field Theories,}
Phys.\ Lett.\ B {\bf 222}, 419 (1989)\semi
F.~Delduc, N.~Maggiore, O.~Piguet and S.~Wolf,
{\it Note on constrained cohomology,}
Phys.\ Lett.\ B {\bf 385}, 132 (1996)
[arXiv:hep-th/9605158]\semi
F.~Fucito, A.~Tanzini, L.~C.~Vilar, O.~S.~Ventura, C.~A.~Sasaki and S.~P.~Sorella,
{\it Algebraic renormalization: Perturbative twisted considerations on 
topological Yang-Mills theory and on N = 2 supersymmetric gauge
theories,} arXiv:hep-th/9707209.}

%%%%%%%%%%%%%%%%%%%%%%%%%%%%%%%%%%%%%%%%%%%%%%%%%%
\baselineskip14pt

{\it Introduction.}~~Recently, we developed a new approach to 
the long-standing problem of the covariant quantization of the superstring \GrassiUG. 
The formulation of  Berkovits of super-Poincar\'e covariant superstrings 
in $9+1$ dimensions \berko~ is based on a {\it free} conformal field theory 
on the world-sheet and a
nilpotent BRST charge which defines the physical vertices as
representatives of its cohomology. In addition to the conventional
variables $x^m$ and $\t^\a$ of the Green-Schwarz formalism, a
conjugate momentum $p_\a$ for $\t^\a$ and a set of commuting ghost fields
$\l^\a$ are introduced. The latter are complex Weyl spinors satisfying the
pure spinor conditions $\l^\a \g^m_{\a\b} \l^\b = 0$ (cf. for example \howe). 
This equation can be solved by decomposing $\lambda$ with
respect to a non-compact $U(5)$ subgroup of $SO(9,1)$ into a singlet
$\underline{1}$, a vector $\underline{5}$, and a tensor
$\underline{10}$. The vector can be expressed in terms of the singlet
and tensor, hence there are 11 independent complex variables in
$\l^\a$.

Since the presence of the non-linear constraint $\l^\a \g^m_{\a\b}\l^\b = 0$ 
makes the theory unsuitable for a path integral
quantization and higher loop computations, we relaxed the pure spinor
condition by adding further ghosts. We were naturally led to a finite
set of extra fields, but the BRST charge $Q$ of this system was not
nilpotent, and the central charge of the conformal field theory did
not vanish. The latter problem was solved by adding one more extra
ghost system, which we denoted by $\eta^m$ and $\omega^m_z$. The
former problem was solved by introducing yet another new ghost pair,
$b$ and $c_z$, which we tentatively associated with the central charge
generator in the affine superalgebra which plays an essential role in the
superstring \csm.

The BRST charge is linear in $c_z$, and without further conditions on physical states the theory 
would be trivial. We proposed that physical states belong not only to the BRST cohomology
($ Q \, |\psi \rangle = 0$, but $ |\psi \rangle \neq Q\, |\phi\rangle$), but also that 
the deformed stress tensor $T+ {\cal V}^{(0)}$, where ${\cal V}^{(0)}$ is a vertex operator, 
satisfies the usual OPE of a conformal spin 2 tensor. (The latter condition is weaker that the 
requirement that vertex operators be primary fields with conformal spin equal to 1).  

In this letter we propose another definition of physical states. We retain 
the BRST condition, but we replace the stress tensor condition by the 
requirement that the physical states belong to a subspace ${\cal H}'$ 
of  the entire linear space ${\cal H}$ of vertex operator. The latter can be decomposed 
w.r.t. a grading naturally associated with the underlying affine algebra (cf. \kacmoody)
as ${\cal H} = {\cal H}_- \oplus {\cal H}_+ $, with negative and 
non-negative grading, respectively. The  BRST charge $Q = \sum_{n\geq 0} Q_n $ contains only 
terms $Q_n$ with non-negative grading, hence one can consistently consider the action of 
$Q$ in ${\cal H}_+$. The physical 
space is identified with the cohomology group  $H(Q, {\cal H}_+)$, 
namely 
\eqn\leb{\eqalign{ &  Q |\psi \rangle = 0\,, ~~~~~~~~~~ 
|\psi \rangle~ \in {\cal H}_+ \,, \cr
& |\psi \rangle \neq Q\, |\phi \rangle\,, ~~~~~~ |\phi \rangle~ \in {\cal H}_+\,.
}}

We show that with these conditions we obtain in a very simple way the correct massless 
spectrum of the open and closed superstring. In particular, our results agree with 
the work of Berkovits and collaborators \berko, 
who obtains the correct field equations for the massless states
$\g^{\a\b}_{mnrpq} D_\a A_\b=0$, but defines physical 
states by only the BRST condition on the unintegrated vertex ${\cal U}^{(1)} = \l^\a A_\a (x,\t)$ 
in terms of pure spinors $\l^\a$. In their work the superfields $A_m$ and 
$W^\a$ (the latter is denoted by $A^\a$ in \GrassiUG) of the open string are expressed 
in term of $A_\a$ by the relations of $d=10$ superspace, 
\eqn\lea{
A_m = {1\over 8} \g_m^{\a\b} D_\a A_\b\,, ~~~~~~~~~~
W^\a \equiv A^\a = {1\over 10} \g^{m,\a\b} \Big( D_\b A_m - \p_m A_\b \Big)\,,
}
whereas in our work these same relations follow 
from the physical state conditions. We obtain corresponding results for closed 
(type II) strings, which also agree with \berko~and \BerkovitsUE. 

Then we consider the sectors with ghost number different
from one,and obtain in all these sectors the correct results.

We also give a derivation of the $\eta^m, \omega^m_z$ terms in the action, based on a 
local symmetry which has been found in the constrained spinor approach \BerkovitsRB. 
This derivation leads us to replace the pair $(\eta^m, \omega^m_z)$ in \GrassiUG~with 
$(\eta^m_z, \omega^m)$. 
%At the end we will comment on the role of the $b,c_z$ system in 
%this approach. 

{\it The new current and grading.}~~We based our approach on the 
following affine superalgebra \csm 
\eqn\dope{\eqalign{
&d_\a(z) d_\b(w) \sim -{{\g^m_{\a\b}\Pi_m(w)}\over{z-w}},\quad \quad\quad\quad
d_\a(z) \Pi^m(w) \sim{{\g^m_{\a\b}\p\t^\b(w)} \over {z-w}}, \cr
&\Pi^m(z) \Pi^n(w) \sim- {1\over (z-w)^2} \, \eta^{mn} \, k \,, \quad ~
d_\a(z) \p_w \t^\b(w) \sim{1\over (z-w)^2} \, \delta^{~\b}_{\a}\, k\,, \cr 
& \Pi^m(z) \p_w \t^\b(w) \sim 0\,, ~~~~~~~~~~~~~~~~\quad\quad 
\p_z \t^\a(z) \p_w \t^\b(w) \sim 0\,,
}} 
where $\sim$ denotes the singular contributions to the OPE's. 
%The operators $d_\a$, $Pi^m$ and $\p_z \t^\a$ are invariant 
%under supersymmetry. 
This algebra has a natural grading defined as follows: 
$d_\a(z)$ has grading $1/2$, $\Pi^m(z)$ has grading $1$, 
$\p_z \theta^\a(z)$ has grading $3/2$, and the central charge $k$ (which 
numerically is equal to unity) has grading $2$. The corresponding ghost systems are 
$(\l^\a, \b_{z \a})$, $(\xi^m, \b_{z m})$, $(\chi_\a, \kappa^{\a}_z)$, and $(c_z, b)$. We thus 
define the following grading for the ghosts and corresponding antighosts 
\eqn\gr{\eqalign{
& {\rm gr}(\l^\a) = {1\over 2}\,,~~ \quad {\rm gr}(\xi^m) = {1}\,, \quad
~~~{\rm gr}(\chi_\a) = {3\over 2}\,,~~ \quad {\rm gr}(c_z) = {2}\,, \cr
& {\rm gr}(\b_\a) = -{1\over 2}\,, \quad {\rm gr}(\b_m) = {-1}\,, \quad
{\rm gr}(\kappa^\a) = -{3\over 2}\,, \quad {\rm gr}(b) =-{2}\,.
}} 
We also need the ghost $\omega^m$ and the antighost $\eta^m_z$, 
although this pair does not seem to correspond to a generator. 
We assign the grading 
${\rm gr}(\eta^m_z) = - 2$ and ${\rm gr}(\omega^m) = 2$ for the following 
reason. In \GrassiUG, we relaxed the pure spinor constraint by 
successively adding quartets starting from $(\l_+, \l_{[ab]}; \b^+, \b^{[ab]})$ of 
\berko~(the indices $a,b$ belong to the fundamental representation of the $U(5)$ 
subgroup of $SO(1,9)$), and adding the fields 
$(\l^a, \b_a; \xi^a, \b'_a)$ with grading $(1/2,-1/2,1,-1)$. This procedure yields the covariant 
spinors $\l^\a$ and $\b_{\a}$, but now the fields $(\xi^a, \b'_a)$ are 
non-covariant w.r.t. $SO(9,1)$. Thus, we added the quartet 
$(\xi_a, \b^{'a}; \chi_a, \kappa^a)$ with grading  $(1,-1, 3/2,-3/2)$. The spinors 
$(\chi_a, \kappa^a)$ are part of a covariant spinor and the missing parts are 
introduced by adding the quartets $(\chi^+, \kappa_+, c, b)$  
and $(\chi^{[ab]}, \kappa_{[ab]}, \eta^m, \omega^m)$, both with 
grading $(3/2,-3/2, 2,-2)$. In this way, we obtain the covariant fields $\l^\a = (\l_+, \l^a, \l_{[ab]})$; 
$\b_\a = (\b^+, \b_a, \b^{[ab]})$; $\xi^m = (\xi^a, \xi_a)$; $\b^m = (\b^{'a}, \b'_a)$; 
$\chi_\a =(\chi^+, \chi_a, \chi^{[ab]})$; $\kappa^\a= (\kappa_+, \kappa^a, \kappa_{ab})$; 
$b,c$ and $\eta_{m} ,\omega^m$. 

As usual for a conformal field theory, it is natural to introduce a current 
whose OPE's with the ghost and antighosts reproduce the grading assignments in \gr
\eqn\currgrad{
j_z^{grad} = - {1\over 2} \b_{z,\a} \l^\a - \b_{z,m} \xi^m - {3\over 2} \k^\a_{z} \chi_\a - 2 \, b\, c_z 
- 2 \, \eta^m_z \omega_{m}\,.
}
Independent confirmation that this current might be important 
comes from the cancellation of the anomaly (namely the terms with $(z-w)^{-3}$) in 
the OPE of the stress energy tensor $T_{zz}(z)$ 
(cf. eqs.~(1-3) of ref.~\GrassiUG) with $j_z^{grad}$. In fact, one finds 
\eqn\graano{
c^{grad} =   {1\over 2}\times ( + 16)_{\l\b} + 1 \times (-10)_{\xi\b} +  
{3\over 2} \times (+16)_{\kappa\chi} +
2\times (-1)_{bc} + 2 \times (-10)_{\eta\omega} = 0\,.
}
The requirement that the vertex operators contain only terms with non-negative grading 
will lead to the correct massless spectrum. It will also 
severely restrict the contribution of the vertex operators to correlation functions 
(in the usual RNS approach ghost insertions are needed to compensate the anomaly in the ghost 
current, whereas here we anticipate to need insertions of fields in  
${\cal H}_-$ to compensate the 
non-negative grading of vertex operators ${\cal U}^{(1)} \, \in {\cal H}_+$). 

All the terms in the stress tensor $T_{zz}(z)$ and in the ghost current 
\eqn\ghostfinal{\eqalign
{
T_{zz} &= -\half \Pi^m_z  \Pi_{m z} - d_{z \a} \p_z \t^\a - \b_{z m} \p_z \xi^m - 
\b_{z \a} \p_z \l^\a - \k^\a_z \p_z \chi_\a + \p_z b\, c_z -  
\eta^m_z  \p_z \, \omega_m \,, \cr
J^{gh}_z &= - \left( 
\beta_{m z} \xi^m + \kappa_z^\a \chi_\a + \beta_{z \a} \l^\a + b\, c_z 
+ \eta^m_z \omega_{m} \right)   \,,
}}
have grading zero, since they are sums of terms of ghost and antighost pairs with 
opposite grading. On the other hand, the terms in the 
 current $j^B_z(z)$ (cf. eq.~(1.2) in \GrassiUG) and the field $B_{zz}(z)$ 
have different grading\foot{In \GrassiUG~we presented four different solutions $B^{i}$ of the 
the equation $T_{zz}(z) = \{ Q, B^{i}_{zz}(z)\}$. None of the solutions $B^{i}$ have definite 
grading except $B^{IV}_{zz}(z)  = 
b \, \hat{T}_{zz}(z)  + b\p_z b c_z - \half \p^2_z b$ which has grading equal to 
$-2$ carried by the antighost $b$.}. For instance, the BRST current can be 
decomposed in the following pieces $j^B_z(z) = \sum_{n=0}^2 j^{B,(n)}_z(z)$
\eqn\decBRST{\eqalign{
j^{B,(0)}_z(z) &= -\, \xi^m \kappa^\a_z \g_{m \a\b} \l^\b 
-  {1\over 2} \l^\a \g^m_{\a\b} \l^\b \b_{z m} + \cr
& - \half b \left( \xi^m \p_z \xi_m - {3\over 2} \chi_\a \p_z \l^\a + \half \p_z \chi_\a \l^\a \right) 
- {1\over 2} \p_z \left( b\, \chi_\a \l^\a \right) \,, \cr
  j^{B,({1 \over 2})}_z(z) &= \l^\a d_{z \a} \,, 
~~~~~~~~~~~~~~~j^{B,({1})}_z(z) = - \xi^m \Pi_{z m} \,, \cr
 j^{B,({3 \over 2})}_z(z) &= - \chi_\a \p_z \theta^\a \,, ~~~~~~~~~~~j^{B,({2})}_z(z) = c_z \,.      
}} 
It is clear that all terms in $j^B_z(z)$ have non-negative grading. 

{\it Massless Spectrum of the Open Superstring.}~~We now turn to the determination 
of the massless cohomology for the open superstring, namely we will compute $H(Q,{\cal H}_+)$ 
for any ghost number.  

Following the RNS approach, the cohomology $H^{(1)}(Q,{\cal H}_+)$ 
at ghost number $1$ should be identified with the physical fields of the 
open superstring. In particular, 
the massless spectrum coincides with spin zero world-sheet fields ${\cal U}^{(1)}(z)$.  
The most general scalar expression in ${\cal H}_+$ is 
\eqn\cohoC{\eqalign{
{\cal U}^{(1)}(z) &= \l^\a A_\a + \xi^m A_m + \chi_\a W^\a  + \omega^m B_m \cr
& + b\, \Big( \xi^m \xi^n F_{m n}   
+ \l^\a \chi_\b F^{~~\b}_\a + \chi_\a \, \xi^m F^{\a}_{~~m} + \chi_\a \chi_\b F^{\a \b} \Big) \cr 
& + b\, \omega^m \Big( \lambda^\a G_{m\a} + \xi^n G_{mn} + \chi_\a G_{m}^{~~\a} \Big) + 
b\, \omega^m \omega^n K_{mn} \,, 
}}
where $ A_\a, \dots, K_{mn}$ are arbitrary superfields of $x_m, \t^\a$. It clearly makes 
a big difference that $\omega^m$ no longer is a $1$-form. Notice that 
the requirement that ${\cal U}^{(1)}(z)  \in {\cal H}_+$ forbids the terms 
$ b \, (\l^\a \l^\b F_{\a\b} + \l^\a \xi^m F_{\a m})$ in the vertex (these 
terms are indeed present in eq.~(6.4) of \GrassiUG).  

The condition $\{ Q, {\cal U}^{(1)}(z) \} = 0$ implies the following equations 
\eqn\cohoD{\eqalign{ 
%\l\l:~~
& D_{(\a} A_{\b)} - \half \g^m_{\a\b} A_m = 0 \,, \cr 
%\l\xi:~~
& \p_m A_\a - D_\a A_m + \g_{m\a\b} W^\b = 0 \,, \cr 
% \xi\xi:~~
& \p_{[m} A_{n]} + F_{mn} = 0 \,,  \quad\quad %\l\chi:~~ 
D_\b W^\a + F_\b^{~~\a} = 0 \,, \cr
%\xi\chi:~~
& \p_m W^\a + F_{~~m}^{\a} = 0 \,, \quad\quad %\chi\chi:~~ 
F^{\a\b} = 0 \,, \cr
%\l\omega:~~
& D_\a B_m - G_{m\a} = 0\,, \quad \quad %\xi\omega:~~ 
\p_n B_m - G_{mn} = 0 \,, \cr
%\chi\omega:~~
& G_m^{~~\a} = 0\,, \quad \quad \quad \quad 
%\omega\omega: 
\quad \quad K_{mn} = 0\,,
}}
where $D_\a \equiv \p / \p \t^\a + \half \t^\b \g^m_{\a\b} \p / \p x^m$
\foot{The normalization is chosen 
such that $D_\a D_\b + D_\b D_\a = \g^m_{\a\b} \p_m$. We define $D_{(\a} A_{\b)} = 
\half \left( D_\a A_\b + D_\b A_\a \right)$ and $\p_{[m} A_{n]} = 
\half \left( \p_m A_n - \p_n A_m \right)$.}.
The terms in  $\{ Q, {\cal U}^{(1)}(z) \} $ which contain the field $b$ yield 
equations which are the Bianchi identities \GrassiUG. 
From the first two equations of \cohoD~one gets the field 
equations for $N=1, d=(9,1)$ SQED  
\wittwi\ 
\eqn\neh{\eqalign{
& \g_{[mnpqr]}^{\a\b} D_\a A_\b = 0 \,, 
\cr}} 
as well as the definition of the vector potential $A_m$ and 
the spinorial field strength $W^\a$ in terms of $A_\a$
\eqn\nehB{\eqalign{
A_m = {1\over 8} \g_m^{\a\b} D_\a A_\b \,, \quad \quad
W^\a = {1\over 10} \g_m^{\a\b} \left( D_\b A_m - \p_m A_\b \right)\,. }   
}
Moreover, the remaining equations in \cohoD~ imply that the curvatures 
$F_{mn}, F^{\a}_{~~m}$, and $F_{\b}^{~~\a}$ are expressed in terms of the 
spinor potential $A_\a$, and similarly the superfields $G_{m\a}, G_{mn}$ are 
solely functions of $B_m$. 

The gauge transformations 
of the vertex ${\cal U}^{(1)}(z)$ are generated by the BRST variation of 
spin zero field $\Omega^{(0)}(z) \in {\cal H}_+$ with ghost number zero, whose most 
general expression is given by  $\Omega^{(0)}(z) = C + b\, \omega^m M_m$, 
with  $C $ and $M_m$ arbitrary superfields. The BRST variation of   $\Omega^{(0)}$ is  
\eqn\tra{
\delta {\cal U}^{(1)}(z) = \Big[ Q, \Omega^{(0)}(z) \Big] = \l^\a D_\a C 
+ \xi^m \p_m C + \omega^m M_m + 
b \, \omega^m (\l^\a D_\a M_m + \xi^n \p_n M_m ) \,.
} 
One can easily check that the field $M_m$ can be
used to gauge away $B_m$, while $C$ is the usual parameter of the gauge transformations on 
the SQED potentials: $\delta A_\a = D_\a C, ~~\delta A_m = \p_m C$. Thus, the only independent 
superfield is $A_\a$, and it satisfies \neh~which is gauge invariant. 

In order to exhibit explicitly the physical degrees of freedom 
it is convenient to work in the gauge $\theta^\a A_\a =0$. 
The photon $a_m$ and the photino $u^\a$ are identified with the first coefficients 
in the expansion of $A_m$ and $W^\a$ in powers of $\theta$
\eqn\decAA{\eqalign{
A_m(x, \t) &= a_m(x) +  \t^\a \g_{m \a\b} u^\b(x) + \dots \,, \cr
W^\a(x, \t) &= u^\a(x) +  \g^{mn,\a}_{~~~~~~\b} \t^\b \p_m a_n(x) + \dots \,.
}}
The ellipses denote terms which contain derivatives of $a_m$ and $u^\a$. From 
the Bianchi identities and \neh~one derives $\p_m F^{mn} = 0$ and $\g^m_{\a\b} \p_m W^\b =0$, 
while $\t^\a A_\a=0$ implies that 
$A_\a = \half a^m (\g_m \t)_\a + a^{m_1\dots m_5} (\g_{m_1\dots m_5} \t)_\a 
+ (\g^m \t)_\a (u \g_m \t) + (\g_{m_1\dots m_5} \t)_\a v^{m_1\dots m_5}_\b \t^\b + \dots$. 
From \neh~it follows that $a^{m_1\dots m_5} =v^{m_1\dots m_5}_\b = 0$ 
and in this way one has obtained the linearized field 
equations for the gauge field $a_m$ and the gaugino $u^\a$. 

At other ghost numbers the cohomology groups $H^{(n)}(Q, {\cal H}_+)$ 
vanish, except the group $H^{(2)}(Q, {\cal H}_+)$ which contains the antifields 
$a^*_m, u^{*}_\a$ of the gauge field $a_m$ and of the gaugino $u^\a$ \BerkovitsRB.  
The analysis can be performed along the lines of the previous discussion and one 
can see that all the $\omega$-dependent terms are cohomologically trivial, and therefore can 
be reabsorbed by a gauge transformation. For the $\omega$-independent terms one has 
the following general decomposition 
\eqn\decdue{\eqalign{
{\cal U}^{(2)} 
=& \l^\a \l^\b A^*_{\a\b} + \l^\a \xi^m A^*_{\a m} + \dots + \chi_\a \chi_\b A^{*,\a\b} \cr 
+& b\, \Big( \l^\a \l^\b \xi^m F^*_{\a\b m} +  \l^\a \xi^m \xi^n F^*_{\a m n} + 
  \l^\a \l^\b \chi_\g F_{\a\b}^{*~\g} + \dots + \chi_\a \chi_\b \chi_\g F^{*,\a\b\g}\Big)\,. 
}}
The condition ${\cal U}^{(2)}  \in {\cal H}_+$ does not allow the term 
$ \l^\a \l^\b \l^\g F^*_{\a\b\g}$, which is permitted by ghost number counting. 
This implies that the superfield $A^*_{\a\b}$ should satisfy the field equation 
$D_{(\a} A^*_{\b\g)} - {1\over 2} \g^m_{(\a\b}  A^*_{\g) m} = 0$. Furthermore, 
the gauge transformations are generated by the BRST variation of 
a ghost-number $1$ superfield $\Omega^{(1)}(z)$ (whose decomposition is given 
in eq.~\cohoC), namely $\delta {\cal U}^{(2)}(z) \equiv \{ Q, \Omega^{(1)}(z) \}$. One obtains 
that the gauge transformations of the super-antifield $A^*_{\a\b}$ are the 
equations of motion of the potential $A_\a$
\eqn\gaudue{
\delta A^*_{\a\b} = D_{(\a} A_{\b)} - \half \g^m_{\a\b} A_m  \,.
}
As shown in \berko~the only states in the BRST cohomology at non-zero 
momentum are the on-shell gauge field $a_m$ and the gluino $u^\a$ at ghost number $+1$ and 
the corresponding antifields $a^*_m, u^{*}_{\a}$ at ghost number $+2$. The 
latter are the coefficients of the superfield $A^*_{\a\b}$. In fact, since 
$A^*_{\a\b}$ is a symmetric bispinor, it can be decomposed into a vector part and a 
self-dual $5$-form one: $A^*_{\a\b} = \g^m_{\a\b} A^*_m + 
\g^{[mnrpq]}_{\a\b} A^*_{[mnrpq]}$. 
The gauge transformations \gaudue~remove the vector component 
$A^*_m$, and the first coefficients of the $\theta$-expansion of $A^*_{[mnrpq]}$  
exhibit the on-shell fields
\eqn\decanti{
A^*_{[mnpqr]} = (\t \g_{[mnp}\t) (\t \g_{qr]})^\a  u^{*}_{\a}(x) + 
(\t \g_{[mnp}\t) (\t \g_{qr]s})^\a  a^{*,m}(x) + \dots\,. 
}
The ellipses involve terms with derivatives of $a^*_m$ and $u^{*}_{\a}$.

Before closing this section, we point out that the choice of the 
subspace ${\cal H}_+$ is one of the possible choices. Another interesting 
situation is the restriction to the subspace ${\cal H}'_+$ with strictly positive grading. 
It is straightforward to see that, at ghost number 
1, the cohomology $H^{(1)}(Q, {\cal H}'_+)$ is identified with the gauge potentials 
which satisfy $F_{mn} = D_\a W^\b= 0$. This corresponds to a topological gauge 
model in the target space. 

{\it The Massless Spectrum of the Closed Superstring.}~~According to our 
formalism \GrassiUG, the BRST charge is the sum of the BRST charge for the left and 
right movers denoted by $Q_L$ and $Q_R$, respectively. In the super-Poincar\'e 
covariant formulation of closed superstring, one can choose the spinors 
$\t_L^\a$ and $\t^\ha_R$ to be Weyl or anti-Weyl in the left- or in the right-mover 
sector: the same chirality for both spinors leads to Type IIB strings, 
opposite chiralities lead to Type IIA strings. In the following we will not distinguish between 
the two cases. 

As in \berko, also in our formalism the holomorphic and anti-holomorphic sector 
are decoupled. Each operator in the left-mover sector has its counterpart in the 
right-mover one.
%All the quantities such as the 
%grading current $j^{grad}_z$ of \currgrad, the 
%ghost current $J^{gh}_z$, and the stress tensor $T_{zz}$  have their 
%homologous in the anti-holomorphic sector. 
Therefore, the assignments 
of ghost number and grading, 
as well as the central charge cancellations (cf. eq.~\graano), 
are exactly the same in both sectors. 

Denoting by ${\cal H}^L_+$ and ${\cal H}^R_+$ the left- and right-mover 
subspaces with non-negative gradings,  the physical state condition is 
expressed by 
\eqn\clophi{\eqalign{
&Q_L |\psi\rangle = Q_R |\psi\rangle = 0\,,~~~~~~~~~~ 
|\psi\rangle \in {\cal H}^L_+\otimes {\cal H}^R_+\,, 
\cr
& |\psi\rangle \neq Q_L |\phi \rangle +  Q_R\, |\chi\rangle\,,~~~~
~~~~ |\phi\rangle, \, |\chi\rangle 
\in {\cal H}^L_+ \otimes {\cal H}^R_+ \,, \cr
& Q_R  \, |\chi\rangle = Q_L |\phi\rangle\ = 0\,. 
}}

The cohomology group $H^{(1,1)}(Q, {\cal H}^L_+\otimes {\cal H}^R_+)$ is identified 
with the physical degrees of freedom. In particular, the spin zero vertex operator  
${\cal U}^{(1,1)}$ contains the massless fields of the closed superstring.
To determine the linearized field equations, we first analyze the $\omega, \ho$-dependent 
terms of the vertex. Since $\omega$ and $\ho$ are BRST inert, one can analyze the 
sectors of $H^{(1,1)}(Q, {\cal H}^L_+\otimes {\cal H}^R_+)$  with 
different powers of $\omega$ and $\ho$ separately. We prove that all the 
sectors with positive powers of $\omega$ and $\ho$ vanish. Consider the 
generic decomposition with one power of $\omega$, 
${\cal U}^{(1,1)} = \omega^m \Big( {\cal B}_m(\hat C) + b\, C^A {\cal F}_{m A}(\hat C)\Big)$, 
where $C^A$ and $\hat C^\hA$ denote collectively 
the ghost fields $(\l^\a,\xi^m,\chi_\a,b)$ of left- and right-moving sectors. 
The functions ${\cal B}_m$ and ${\cal F}_{m A}$ are polynomials 
in the right-moving ghost fields $\hat C^\hA$ with superfield coefficients.  

The conditions $[ Q_L, {\cal U}^{(1,1)}] = [ Q_R, {\cal U}^{(1,1)}] = 0$ imply that 
\eqn\omme{
Q_L \, {\cal B}_m(\hat C)  - C^A {\cal F}_{m A}(\hat C) = 0\,,~~~~~
Q_R \, {\cal B}_m(\hat C) = Q_R  \, {\cal F}_{m A}(\hat C) = 0\,. 
}
Thus, the vertex ${\cal U}^{(1,1)} = \{ Q_L , \Lambda\}$ is BRST trivial 
where $\Lambda =  \Big( b\, \omega^m  {\cal B}_m(\hat C) \Big)$ and 
$\{ Q_R ,\Lambda \} = 0 $. The same argument can be easily repeated for 
any positive power of $\omega$ or $\ho$. Therefore, we have to analyze only 
the $\omega,\ho$-independent terms. 

The vertex ${\cal U}$ can be decomposed into 
the following terms
\eqn\clovert{\eqalign{
{\cal U}(z, \bar z) &= 
{\cal U}^{(1,1)} + b \, {\cal U}^{(2,1)}  + \hat b \, {\cal U}^{(1,2)} + b \, \hat b \, {\cal U}^{(2,2)}\,, \cr
{\cal U}^{(1,1)} &= \Big( \l_L^\a \l^\ha_R A_{\a\ha} + \dots + \chi_{L \a} \chi_{R \ha} A^{\a\ha}
\Big) \,, \cr
{\cal U}^{(2,1)} &= \Big( \xi^m_L \xi^n_L \l^\ha_R B_{m n \ha} + \dots + 
\chi_{L \a} \chi_{L \b} \chi_{R \ha} B^{\a\b\ha} 
\Big)\,, \cr
{\cal U}^{(1,2)} & = \Big(\l^\a_L \xi^{\hat m}_R \xi^{\hat n}_R  B_{\a \hat m \hat n } + \dots + 
\chi_ {\a} \chi_{R \ha} \chi_{R \hat \g} B^{\a\ha\hat \g} 
\Big)\,, \cr
{\cal U}^{(2,2)} &= \Big( \xi^m_L \xi^n_L  \xi^{\hat m}_R \xi^{\hat n}_R  
C_{m n \hat m \hat n } + \dots + 
\chi_{L \a} \chi_{L \b} \chi_{R \hat \g} \chi_{R \hat \e} C^{\a\b\hat \g\hat \e} 
\Big)\,,
}}
where all the coefficients $A_{\a\ha}, \dots C^{\a\b\hat \g\hat \e}$ are arbitrary superfields. 
Notice that the condition ${\cal U}(z, \bar z)  \in {\cal H}^L_+\otimes {\cal H}^R_+$
forbids the terms $b\,  \l_L^\a \l_L^\b, b\,  \l_L^\a \xi_L^m$ and 
the corresponding ones for right-movers. Finally, also the terms 
$b\, \hat b\, \l_L^\a \l_L^\b \l_R^\ha \l_R^\hg, \dots, 
b\, \hat b\,  \l_L^\a \xi_L^m  \l_R^\ha \xi_R^{\hat m}$ are ruled out. 

As in the case of the open superstring, the absence of these 
terms combined with the BRST invariance yields the following 
equations\foot{The N=2 D=9,1 supersymmetric derivatives
are defined by $D_\a = {\p\over{\p\t^\a}} +{1\over 2} \g^m_{\a\b}\t^\b\p_m$ and
$\hat D_\ha = {\p\over{\p\hht^\ha}} +{1\over 2} \g^m_{\ha\hbe}\hht^\hbe\p_m$.}
$${\eqalign{
& D_{(\a} A_{\b) \ha} - \half \g^m_{\a\b} A_{m\ha} = 0 \,, \quad\quad \quad\quad \quad \quad
 D_{(\ha} A_{\a \hg)} - \half \g^m_{\ha\hg} A_{\a m} = 0 \,, \cr
& \p_m A_{\a \ha} - D_\a A_{m \ha} + \g_{m\a\b} A^\b_{~~\ha} = 0 \,,  \quad \quad \,
 \p_m A_{\a \ha}  - D_\ha A_{\a m}  + \g_{m\ha\hg} A_\a^{~~\hg} = 0 \,, \cr
& D_{(\a} A_{\b) m} - \half \g^n_{\a\b} A_{m n} = 0\,,  \quad\quad \quad\quad \quad \quad
D_{(\ha} A_{m \hg)} - \half \g^n_{\ha\hg} A_{m n} = 0\,, %\cr
}}$$
\eqn\cloeqq{\eqalign{
& \p_m A_{\a n} - D_\a A_{m n} + \g_{m \a\b} A^{\b}_{~~n} = 0 \,,  \quad \quad \,\,
\p_m A_{ n \ha} - D_\ha A_{m n} + \g_{m \ha\hg} A^{~~\hg}_{n} = 0 \,, \cr
&D_{(\a} A_{\b)}^{~~\hg} - \half \g^m_{\a\b} A_{m}^{~~\hg} =0 \,,  
\quad\quad \quad\quad \quad \quad \,
D_{(\ha} A^\a_{~~\hg)} - \half \g^m_{\ha\hg} A_{m}^\a = 0\,, \cr
&\p_m A_{\a}^{~\hg} - D_\a A_m^{~\hg} + \g_{m \a\b} A^{\b\hg} = 0 \,, \quad \quad \quad 
\p_m A^{\a}_{~\ha} - D_\ha A_{~m}^{\a} + \g_{m \ha\hg} A^{\a\hg} = 0 \,.
}}
The other equations coming from $[ Q_L + Q_R, {\cal U}] = 0$ 
will not impose any further constraint on the superfields, but they define consistently 
the curvature in terms of $A_{\a\ha}, \dots A^{\a\ha}$. 
Notice that the superfields $A_{m \ha},A_{\a m}, \dots A^{\a\ha}$
are solved in terms of the fundamental field $A_{\a\ha}$ as follows
\eqn\clonehA{\eqalign{
&A_{m \ha} = {1\over 8}D_\a \g_m^{\a\b} A_{\b \ha}\,,\quad\quad\quad\quad\quad\quad
\quad\quad\quad
A_{\a m} = {1\over 8}  D_\ha \g_m^{\ha \hbe} A_{\a \hbe}\,, \cr
& A^\a_{~~\ha} = {1\over 10} \g^{m\a\b} \Big(   D_\b A_{m \ha} - \p_m A_{\b \ha}\Big)\,, 
\quad\quad
 A_\a^{~~\ha} = {1\over 10} \g^{m\ha\hg} \Big(   D_\hg A_{\a m } - \p_m A_{\a \hg}\Big)\,, \cr
& A_m^{~~\hg} = {1\over 8} \g^{\a\b}_m 
D_\a A_{\b}^{~~\hg}\,, \quad\quad\quad\quad\quad\quad
\quad\quad
 A^{\a}_{~~m} = {1\over 8} \g_m^{\ha\hg} D_\ha A^{\a}_{~~\hg}\,, \cr
&A_{m n} = {1\over {64}}D_\a  D_\ha \g_m^{\a\b} \g_n^{\ha\hbe} A_{\b \hbe}\,, 
\quad\quad\quad
\quad\quad\,\,
A^{\a \ha} = {1\over 10} 
\g^{m\a\b} \Big( D_\b A_{ m}^{~~\ha} - \p_m A_{\b}^{~~\ha}\Big)\,. \cr
}}
The superfield $A_{\a\ha}$ satisfies the field equations
\eqn\cloneh{
\g_{mnpqr}^{\a\b} D_\a A_{\b \hg} = 0\,,\quad
\g_{mnpqr}^{\ha\hg}  D_\ha A_{\g \hg} = 0\,,
}
which are the linearized $N=2$ supergravity equations of motion. 
The gauge transformations are generated by the BRST variations 
of two generic superfields $\Omega^{(0,1)}$ and $\Omega^{(1,0)}$ by 
$\delta {\cal U}(z, \bar z)  = \{ Q_L, \Omega^{(0,1)} \} +  \{ Q_R, \Omega^{(1,0)} \}$, 
with $\{ Q_R, \Omega^{(0,1)} \} = \{ Q_L, \Omega^{(1,0)} \} = 0$ and 
$\Omega^{(0,1)} \in {\cal H}_+, \, \Omega^{(1,0)} \in {\cal H}_+$. 
Assuming for $\Omega^{(0,1)}$ the general decomposition 
\eqn\cohocloseC{\eqalign{
\Omega^{(0,1)}(z) &= \l^\ha_R \hat\Lambda_\ha + \xi^m_R \hat\Lambda_m + 
\chi_{R\ha} \hat\Lambda^\ha  \cr
& + b_R\, \Big( \xi^m_R \xi^n_R \hat\Xi_{m n}   
+ \l^\ha_R \chi_{R\hg} \hat\Xi^{~~\hg}_\ha + 
\chi_{R\ha} \, \xi^m \hat\Xi^{\ha}_{~~m} + \chi_{R\ha} \chi_{R\hg} \hat\Xi^{\ha \hg} \Big) \,,
}}
and analogously for $\Omega^{(1,0)}$ exchanging  
$R \rightarrow L$ and $\hat\Lambda_\ha, \dots,  \hat\Xi^{\ha \hg} \rightarrow  
\Lambda_\a, \dots,  \Xi^{\a \g}$, 
the relevant gauge transformations are 
\eqn\gaugeclo{\eqalign{
&\delta A_{\a\ha} = D_\a \hat\Lambda_\ha + D_\ha \Lambda_\a\,. 
}}
The conditions $\{ Q_R, \Omega^{(0,1)} \} = 
\{ Q_L, \Omega^{(1,0)} \} = 0$ imply that the gauge parameters 
$\Lambda_\b$ and $\hat\Lambda_\hg$ satisfy the following equations
\eqn\gaugeclocon{\eqalign{
& \g_{[mnpqr]}^{\a\b} D_\a \Lambda_\b = 0 \,,~~~~~~  
\g_{[mnpqr]}^{\ha\hg} D_\ha \hat\Lambda_\hg = 0 \,. 
\cr}} 
The gauge transformations for the other superfields $A_{m\ha}, \dots, A^{\a\ha}$ can 
be easily derived from eq.~\gaugeclocon~using their definitions \clonehA. 
The field equations \cloneh~and the gauge transformations \gaugeclo~derived for the 
closed superstring are in agreement with \berko~and \BerkovitsUE. 
The physical degrees of freedom can be easily read from the 
first components of the $\theta^\a_L$ and $\theta^\ha_R$ of the superfields 
$A_{mn}(x,\t_L,\t_R),  A_m^{~~\hg}(x,\t_L,\t_R), A^\a_{~~m}(x,\t_L,\t_R)$ and 
$A^{\a\ha}(x,\t_L,\t_R)$. The first component of 
$A_{mn}= (g_{mn} + b_{mn} + \eta_{mn} \phi) + {\cal O}(\t_L,\t_R)$ describe the graviton, the 
NS-NS two-form and the dilaton. The first components of 
$A_m^{~~\hg}=\psi_m^{~~\hg} + {\cal O}(\t_L,\t_R)$ and $A^\a_{~~m}= 
\psi_{~~m}^{\a} + {\cal O}(\t_L,\t_R)$ contain the N=2 gravitinos 
$\psi_m^{~~\hg}, \psi_{~~m}^{\a}$, and the 
chirality of the right-mover spinor $\t_R$ determines if the string is Type IIA or 
IIB. Finally, $A^{\a\ha}= F^{\a\ha} + {\cal O}(\t_L,\t_R)$  where $ F^{\a\ha}$ are the 
R-R field strenghts. The linearized field equations for 
$g_{mn}, b_{mn}, \phi, \psi_{~~m}^{\a}, \psi_m^{~~\hg}$ and $F^{\a\ha}$ are 
discussed in great detail in \BerkovitsUE.

Of course the closed superstring massless spectrum  
can be understood as the tensor product of the cohomologies for 
open superstring of left- and right-movers. Mathematically, 
this is encoded in the well-known K\"unneth formula 
$H^{(1,1)}(Q_L+Q_R, {\cal H}^L_+\otimes {\cal H}^R_+) = 
H^{(1)}(Q_L, {\cal H}^L_+) \otimes H^{(1)}(Q_R, {\cal H}^R_+)$.

{\it Conclusions.}~~We conclude by giving an argument for the presence of the term 
$- \eta^m_z \bar\p \omega_m$ in the action, with $(\eta^m_z, \omega_m)$ a ghost-charge 
$(-1,1)$ spin $(1,0)$ system. We start from the observation that in the  approach with the 
pure spinor constraint $\l^\a \g^m_{\a\b} \l^\b = 0$ the action with 
${\cal L} = - \beta_{z,\a} \bar\p \lambda^\a$ has the local gauge invariance 
\eqn\lga{
\delta \b_{z,\a} = \Lambda_{z m} \g^m_{\a\b} \l^\b\,,~~~~ ~~~ \d \l^\a = 0\,. 
}
The gauge parameter has ghost number $-2$. To remove the pure spinor constraint
we introduce the Lagrange multiplier field $\a_{z\bar z, m}$ with the ghost number $-2$
\eqn\lgaA{
{\cal L} = - \beta_{z,\a} \bar\p \lambda^\a + {1\over 2} \a_{z\bar z, m} \l^\a \g^m_{\a\b} \l^\b\,. 
}
The action with unconstrained $\l^\a$ is still gauge invariant if $\a_{z\bar z, m}$ 
transforms as $\d \a_{z\bar z, m} = - \bar\p \Lambda_{z m}$. We fix this local 
gauge symmetry as usual by adding a BRST exact term. The gauge parameter 
$\Lambda_{z m}$ becomes a field $\eta_{z m}$ with ghost number $-1$, which is an antighost. 
As the BRST exact term we take
\eqn\lgaB{
s\, \Big( \omega^m \a_{z\bar z, m} \Big) = d^m \a_{z\bar z, m} - \omega^m \bar\p \eta_{z m}\,.
}
The field $\omega^m$ 
has ghost number $+1$, and is a ghost field; it transforms into a BRST auxiliary field 
$d^m$ with ghost number $+2$. Integrating over $d^m$ and $\a_{z\bar z, m}$  sets these fields 
to zero, and one is indeed left with the term 
$\bar\p \eta_{z m} \omega^m = - \eta_{z m} \bar\p \, \omega^m$ in the action.  
We end up by recalling that the previous argument is a strong indication in favour of 
the presence of the fields $\eta^m_z$ and $\omega_m$  in our formalism, nevertheless 
a complete picture can be only achieved by starting from a gauge invariant action and using the 
Batalin-Vilkovisky framework for its quantization.  

Before concluding, we would like to point out the relation between 
the cohomology in the Berkovits' approach and in our formalism which 
also indicates the path toward the proof of the equivalence of the two approaches. 

The group $H(Q_B|{\cal H}_{p.s.})$, which represents the cohomology in the 
Berkovits' formalism is an example of a constrained BRST cohomology, or 
equivalently, of equivariant cohomology \equivariant. In the latter case, the BRST cohomology 
is computed on the supermanifold $M$ with coordinates $x^m, \theta^\a$ on which the space-time 
translations $ x^m \rightarrow x^m + \half \lambda \g^m \lambda $, generated by 
unconstrained spinors $\l^\a$, act freely. One finds that $Q^2_B = - {\cal L}_{V}$ where 
$V^m= \half \lambda \g^m \lambda$. 
Notice that the r.h.s. can be also written in term of the Lie derivative 
${\cal L}_V = d\, {\iota}_V +   {\iota}_V \, d$, where $  {\iota}_V$ is the 
contraction of a form with the vector $V^m$. One can represent $\iota_V$ by the 
operator $\oint dz \, V^m \b_{zm}$; its action of (parity reversed forms) $\xi^m$ is then 
given by the OPE of $\b_{zm}(z)$ with $\xi^m(z)$. 
The exterior differential $d$ is $\xi^m \p_m$ where $\xi^m$ 
are the parity-reversed coordinates of the cotagent bundle $\Pi T^*{\cal M}$. 
The usual exterior derivative $d = dx^m \p_m $ has been replaced by 
$- \oint dz\, \xi^m \Pi_{zm}$ . Since $\Pi^m_z(z) \p^l x^n(w) \sim (z-w)^{-l-1}$, the 
operator $- \oint dz\, \xi^m \Pi_{zm}$ represents the exterior derivative on the jet bundle 
$\{x^m, \p\, x^m, \p^2 x^m, \dots\}$. 
Following the approach of equivariant cohomology \equivariant, 
one can define a new BRST operator $Q'$ by 
\eqn\newBRST{
Q' = Q_B + d + \iota_V = 
 Q -  \oint \xi^m \Pi_{z m} - \oint {1\over 2} \l^\a \g^m_{\a\b} \l^\b \b_{z m}\,.
}
Unfortunately, this operator fails to be nilpotent and the solution of this problem 
has been discussed in \GrassiUG. Moreover, it turns out that the  
BRST cohomology computed in the space of zero conformal-weight 
vertex operators with non-negative gradings coincides with the massless 
spectrum of the superstring, and this suggests a complete equivalence of the 
two approaches. It has been also pointed out in similar examples discussed in 
\brscohotopo~that one needs further conditions on the functional space to identify 
the correct physical observables as elements of the BRST cohomology. 

In the present paper, we identify the massless spectrum for the open 
and closed superstrings by the cohomology at ghost number $+1$ of the 
BRST operator restricted to a subspace of the entire linear space of vertex operators. 
The subspace is selected by means of a natural grading which is assigned to the 
ghosts and antighosts.  

We are involved in the computation of amplitudes using the present 
formalism. The preliminary results are very encouraging and we 
hope to report on this soon. 

%%%%%%%%%%%%%%%%%%%%%%%%%%%%%%%%%%%%%%%%%%%%%%%%%%%

\vskip 1cm
We thank M. Porrati, N. Berkovits and J.-S. Park for useful discussions. G. P.  thanks 
C.N. Yang Institute for Theoretical Physics at Stony Brook for 
the hospitality.  This work was partly funded by NSF Grants PHY-0098527.
\vfill

\footatend\vfill\supereject\immediate\closeout\rfile\writestoppt
\baselineskip=14pt\centerline{{\bf References}}\bigskip{\frenchspacing%
\parindent=20pt\escapechar=` \input refs.tmp\vfill\eject}\nonfrenchspacing

\bye